\documentclass[aps,prd,showkeys,showpacs,amssymb,eqsecnum,amsfonts,epsf,
preprintnumbers,nofootinbib,superscriptaddress]{revtex4}

\usepackage[dvips]{graphicx}
\usepackage{graphicx}
\usepackage{natbib}
\usepackage{latexsym}
\usepackage{amsfonts}
\usepackage{bm}
\usepackage{url,hyperref}

\begin{document}

\newcommand{\vp}{\varphi}
\newcommand{\nn}{\nonumber\\}
\newcommand{\beq}{\begin{equation}}
\newcommand{\eeq}{\end{equation}}
\newcommand{\bed}{\begin{displaymath}}
\newcommand{\eed}{\end{displaymath}}
\def\bea{\begin{eqnarray}}
\def\eea{\end{eqnarray}}

%%%%%%%%%%%%%%%%%%%%%%%%%%%%%%%%%%
\title{Casimir effect in rugby-ball type flux compactifications}
\author{ Emilio~Elizalde}
\email[Email: ]{elizalde"at"ieec.uab.es}
\affiliation{Consejo Superior de Investigaciones Cient\'{\i}ficas,
ICE/CSIC and IEEC, Campus UAB, Fac Ciencies, 08193 Bellaterra (Barcelona) Spain}
\author{Masato~Minamitsuji}
\email[Email: ]{masato"at"theorie.physik.uni-muenchen.de}
\affiliation{ASC, Physics Department, LMU, Theresienstr. 37, Munich, Germany}
\affiliation{Yukawa Institute for Theoretical Physics, Kyoto University,
Kyoto 606-8502, Japan}
\author{Wade~Naylor}
\email[Email: ]{naylor"at"se.ritsumei.ac.jp}
\affiliation{Department of Physics, Ritsumeikan University,
Kusatsu, Shiga 525-8577, Japan}

\begin{abstract}
As a continuation of the work in \cite{mns},
we discuss the Casimir effect for a massless bulk scalar field
in a 4D toy model of a 6D warped flux compactification model,
to stabilize the volume modulus.
The one-loop effective potential for the volume modulus
has a form similar to the Coleman-Weinberg potential.
The stability of the volume modulus against quantum corrections
is related to an appropriate heat kernel coefficient.
However, to make any physical predictions after volume stabilization,
 knowledge of the derivative of the zeta function, $\zeta'(0)$
(in a conformally related spacetime) is also required.
By adding up the exact mass spectrum
using zeta function regularization,
we present a revised analysis of the effective potential.
Finally, we discuss some physical implications,
especially concerning the degree of the hierarchy between the
fundamental energy scales on the branes.
%%%%%%%%%%%%%%%%%%%%%%%%%%%%%%%%%%%%%%%%%%%%%%%%%%%%%%%%%
For a larger degree of warping our new results are very similar to the ones
given in Ref \cite{mns}
and imply a larger hierarchy.
%%%%%%%%%%%%%%%%%%%%%%%%%%%%%%%%%%%%%%%%%%%%%%%%%%%%%%%%%
In the non-warped (rugby-ball) limit the ratio tends to converge to the same value,
independently of the bulk dilaton coupling.
\end{abstract}

%%%%%%%%Please enter your preprint number%%%%%%%%%%%%%
%\preprint{XXX}
\preprint{LMU-ASC 07/07}
\preprint{RITS-PP-012}
\date{\today}
\maketitle

%%%%%%%%%%%%%%%%%%%%%%%%%%%%
\section{Introduction}

Recent studies on braneworld models with extra dimensions that are compactified by a magnetic flux and bounded by codimension 2 branes (conical singularities) \cite{Sidestep, Gibbons:2003di} have revealed various fundamental properties, i.e., the regularization of branes (conical singularities)/linearized gravity \cite{deRham:2005ci,Peloso:2006cq, Carter:2006uk} and stability against classical perturbations \cite{SusyPerts,nonSUSYpert}. There are still a lot of unsolved questions which should be clarified, especially concerning non-linear gravity and cosmology. Furthermore, careful analysis of these simpler models may give important physical insights on self-gravitating branes in various flux compactifications in string theory \cite{Douglas:2006es}.

In this paper, we focus on models of braneworld based on a
6D supergravity \cite{Gibbons:2003di}, though we will work with its
4D counterpart \cite{4D}, because of the lack of a formulation of the heat kernel coefficients for conical singularities in 6D.
In these supergravity inspired models, the size of the compactified
internal-space is not fixed classically and may behave as a volume
modulus in a 4D effective theory.
To fix this size of the volume (the volume modulus), we should therefore discuss other additional mechanisms.
%%%%%%%%%%%%%%%%%%%%%%%%%%%%%%%%%%%%
The authors of Ref. \cite{mns} focused on Casimir corrections for the case of perturbations of a massless, minimally coupled bulk scalar field.
%%%%%%%%%%%%%%%%%%%%%%%%%%%%%%%%%%%%%%
In this article, we present a revised analysis of the previous
calculations, by performing a strict mode summation of the exact mass
spectrum, which is now wholly taken into account.

We give an exact analysis of the one-loop effective action in a 4D
alternative model.
The one-loop effective potential for the volume modulus can be written
in a form which is somewhat similar to the Coleman-Weinberg potential:
\begin{eqnarray}
V_{\rm 4, eff}
=\frac{A_4-B_4 \ln(\mu^2 \rho_+)}{\rho_+}\,,
\label{effpot}
\end{eqnarray}
where $\rho_+$ characterizes the volume modulus and $A_4$ and $B_4$ are functions of the model parameters and
the shape modulus (the degree of the warping of the bulk, which is completely
determined  if one fixes the brane tensions).
Stability itself is determined by the sign of $B_4$, which is
closely related to the relevant heat kernel coefficients.
However, to discern phenomenological effects on the brane, i.e.,
the effective mass of the modulus and the degree of the hierarchy
between fundamental energy scales on the brane, we need to know
the value of $A_4$ as well.
$A_4$ is not just related to the heat kernel,
although partly related with it; we also need to evaluate
$\zeta'(0)$.
Thus, an accurate evaluation of $A_4$ is crucial to make physical predictions, e.g., for the hierarchy or cosmological constant (CC) problems.

%%%%%%%%%%%%%%%%%%%%%%%%%%%%%%%%%%%%%%%%%%%%%%%%%%%%%%%%%
In the work \cite{mns}, it has been shown how to
divide $A_4$ into two parts, by introducing a continuous conformal transformation,  going from the unwarped frame to the original warped geometry.
%%%%%%%%%%%%%%%%%%%%%%%%%%%%%%%%%%%%%%%%%%%%%%%%%%%%%%%%%
The first part is associated with a one parameter family of conformal transformations and is called the cocycle function.
This term can be obtained from heat kernel analysis.
The second part consists of calculating the derivative of the
zeta function in the unwarped frame, where to
evaluate this piece we used the WKB approximation.
In this paper we shall present an exact analysis of the
mass spectrum for these {\it Kaluza-Klein} like modes in the
unwarped frame.\footnote{These modes are not the standard KK modes,
because of the presence of conical singularities at the poles of the
two-sphere on the internal dimensions. Given the similarity to a rugby
ball we shall also call this unwarped frame the rugby-ball frame.}

The structure of the article is as follows:
In the next section we briefly discuss our 4D analogue
warped flux compactification model and review the tools to analyze the
one-loop Casimir effect.
We give the exact mass spectrum in the unwarped rugby-ball frame.
In Sec.~III we carry out the zeta function regularization in the
unwarped frame.
%%%%%%%%%%%%%%%%%%%%%%%%%%%%%%%%%%%%%%%%%%%%%%%%%%%%%%%
 In Sec.~IV, we give physical implications of our result,
e.g., for the hierarchy problem by
making comparisons with the results in \cite{mns}.
%%%%%%%%%%%%%%%%%%%%%%%%%%%%%%%%%%%%%%%%%%%%%%%%%%%%%%
We finish this article after giving a summary and some discussions.

\section{4D warped flux compactification model}

Our main interest is the Casimir effect in the
warped flux compactification model in 6D
supergravity \cite{Gibbons:2003di}:
\begin{eqnarray}
  S_6=
M_6^4
\int d^6 x\sqrt{-g}
\left(
 \frac{1}{2}R
-\frac{1}{2}\partial_A \varphi \partial^A \varphi
-\frac{1}{4} e^{-\varphi}F_{AB}F^{AB}
 -2g_0^2 e^{\varphi}
\right)\,,
\label{theory6d}
\end{eqnarray}
where $F_{AB}$ is the field strength of the electromagnetic field,
$\varphi$ is the dilaton, and $g_0$ is the bulk dilaton coupling.
Hereafter, we set $M_6^4=1$ and if needed, we put it back explicitly.
In order to evaluate the Casimir energy in this model,
we need to know the heat kernel coefficients on the conical
singularities. However, as mentioned in \cite{mns},
there is no mathematical formulation of these coefficients the
authors are aware of.
Thus, in this paper, we focus on its 4D counterpart theory:
\begin{eqnarray}
  S_4=
M_4^2
\int d^4 x\sqrt{-g}
\left(
  R
-\frac{1}{2}\partial_A \varphi \partial^A \varphi
-\frac{1}{8} e^{-\varphi}F_{AB}F^{AB}
 -4g^2 e^{\varphi}
\right)\,,
\label{theory4d}
\end{eqnarray}
which has a series of warped flux compactification solutions
\cite{4D}
\begin{eqnarray}
ds^2&= & h(\rho)d\theta^2
     +\frac{d\rho^2}{h(\rho)}
     +(2\rho)(-d\tau^2+d x_2^2)\,,
     \quad
      h(\rho)=\frac{2g^2}{\rho}
       \left(\rho_+ -\rho\right)
       \left(\rho -\rho_-\right)\,,
     \nonumber\\
     &&\varphi(\rho)=-\ln(2\rho)\,,\quad
     F_{\theta\rho}=- \frac{ Q}{\rho^2}%\epsilon_{\theta \rho}
\,.
\end{eqnarray}
We also set $M_4=1$ unless it is needed. Then, the branes
correspond to strings, lying at the positions given by the
{\it horizon-like} condition $h(\rho)=0$. It is useful to define the
coordinate $z=((\rho_+ - \rho)/(\rho - \rho_-))^{1/2}$, to resolve
the structure of spacetime in the case that $\rho_+=\rho_-$. For
$z\to 0, \infty$, there are cones whose deficit angles are obtained
from the relation $\delta_{\pm}:= 2\pi-g^2
(\rho_+-\rho_-)\Delta\theta/(\rho_{\pm})$.

The moduli in the effective theory on the branes are characterized by $\rho_{\pm}$, or equivalently by $\rho_+$ and $r$, where the string tensions are related to the conical deficits by $\sigma_{\pm}=  M_4^2 \delta_{\pm}$.
We stress that these relations are only valid for sufficiently small brane tensions, in comparison
with the bulk scale $M_{4}^2$.
The coordinate $\theta$ is defined as $0\leq \theta< \Delta\theta$ and
$\Delta\theta$ is given by
\begin{eqnarray}
 \Delta\theta(r,\delta_+)
=\frac{2\pi-\delta_+}{g^2(1-r)}
      \,,
\end{eqnarray}
where $r:=\rho_-/\rho_+$.
Once the brane tensions, $\sigma_+$ and $\sigma_-$ are fixed,
then $r$ is also fixed and so we now regard as free parameters
$r$ and $\delta_+$, along with the dilaton bulk coupling $g$.
The remaining degree of freedom used to determine the bulk geometry is
the absolute size of the bulk, i.e., $\rho_+$,
which should be fixed by additional mechanisms.

We discuss the basic properties of a massless, minimally coupled scalar field
on this background:\footnote{A related model including a
self-interaction term was discussed in \cite{MOZ}.}
\begin{eqnarray}
S_{\rm scalar}=-\frac{1}{2}
                \int d^4 x\sqrt{g}
                \phi\Delta_4 \phi\,.
    \label{scalaraction4d}
\end{eqnarray}
Now consider a continuous conformal transformation of the
metric, parameterized by $\epsilon$;
\begin{eqnarray}
d{\tilde s}_{4,\epsilon}^2
 = e^{2(\epsilon-1)\omega}ds_4^2\,,
 \qquad
\omega= \frac{1}{2} \ln(2\rho)\,,\label{confort4d}
\end{eqnarray}
where for $\epsilon=1$ we have the original metric, which we shall
define as $\Delta_{4,\epsilon}=\Delta_4$, and $\epsilon=0$ is the
conformal frame.
The classical action changes as
\begin{eqnarray}
  S_{\rm scalar}
   =-\frac{1}{2}\int d^4 x\sqrt{g}
     \phi\Delta_4\phi
   =-\frac{1}{2}\int d^4 x\sqrt{\tilde g}
     \tilde \phi\left(\tilde \Delta_4+E_4(\epsilon) \right)
     \tilde \phi\,,
     \label{class4d}
\end{eqnarray}
where
\begin{eqnarray}
E_4
(\epsilon)
&=& -(\epsilon-1)^2 \tilde{g}^{ab}
                 \nabla_{a} \omega
                  \nabla_{b}\omega
     +(\epsilon-1){\tilde \Delta}_4\ln\omega
   \nonumber\\
 &=&
\left(\frac{1}{2\rho}\right)^{\epsilon}
\frac{g^2(1-\epsilon)(\rho_+ - \rho_-)
   \big\{
     \rho_+ (2+(1-\epsilon)z^2)
+  \rho_- z^2(-1+\epsilon-2z^2)
     \big\} }
     {(\rho_+ + \rho_- z^2)^2}\,.
\end{eqnarray}

We now derive the mass spectrum in the unwarped frame:
\begin{eqnarray}
 \left(
\tilde   \Delta_4
+ E(0)
  \right)
  \tilde \phi_{\lambda} =
-\lambda^2\tilde
\phi_{\lambda}\,,
\label{eigeneq}
\end{eqnarray}
We shall decompose the mass eigenfunction as
\begin{eqnarray}
 \tilde \phi_{\lambda}
=\int
  \frac{d^2k}{(2\pi)}
  \sum_{m,n}
  \Phi_{\lambda}(z)e^{in(2\pi/\Delta\theta)\theta} e^{i {\bf kx }}\,.
\end{eqnarray}
The equation of motion of equation (\ref{eigeneq}) has a series of exact solutions :
\begin{eqnarray}
\Phi_{\lambda}(r)&=&
\sqrt{\frac{1+rz^2}{1+z^2}}
\Big[
 A \left(\frac{z^2}{1+z^2}\right)^{-n /2\kappa}
   \left(\frac{1}{1+z^2}\right)^{ n r/2\kappa}
{}_{2}F_{1}\big(1-\nu-\frac{n}{2\kappa} (1-r),
            \nu-\frac{n}{2\kappa} (1-r),
            1- \frac{n}{\kappa};
           \frac{z^2}{1+z^2}
           \big)
\nonumber\\
&+& B
   \left(\frac{z^2}{1+z^2}\right)^{n  /2\kappa}
   \left(\frac{1}{1+z^2}\right)^{-n r /2\kappa}
{}_{2}F_{1}\big(
           1-\nu+\frac{n}{2\kappa} (1-r),
            \nu+\frac{n}{2\kappa} (1-r),
            1+\frac{n}{\kappa} ;
           \frac{z^2}{1+z^2}
           \big) \label{nonzero}
\Big],
\end{eqnarray}
where $\kappa:= 1-\frac{\delta_+}{2\pi}$.
\begin{eqnarray}
\nu=\frac{1}{2}
 \left(
  1+\sqrt{1+\frac{\lambda^2-k^2}{g^2} +\frac{n^2}{\kappa^2}(1-r)^2}
 \right).
\end{eqnarray}
Here we assume that the mode functions are regular on both conical
boundaries. In order to ensure such a condition at $z=0$ we must have $A=0$ for $n>0$ (and $B=0$ for $n<0$).
Thus, we arrive at the following mass spectrum
\begin{eqnarray}
\lambda^2= k^2 +g^2 \big[4m(m+1)+\frac{2|n|}{\kappa}(2m+1)(1+r)
          +\frac{4n^2r}{\kappa^2}\big],\qquad\qquad
m=0,1,2,\cdots.
\label{eigen}
\end{eqnarray}
Note that this mass spectrum is very similar to the case of
the gauge field perturbations \cite{Parameswaran:2006db}. (See also
\cite{Carter:2006uk} for exact solutions of massless tensor
perturbations).\footnote{Our method to determine the mass spectrum
is essentially based on the same arguments as that in \cite{Parameswaran:2006db}.
They impose two independent conditions, namely normalizability and
regularity (denoted as ``hermitian conditions" in \cite{Parameswaran:2006db}).
However, if we were to only impose the condition of normalizability then there may still be other KK-type modes, though they are somewhat out of the scope of this paper.}

From the above analysis, we can compare the results of the WKB approximation with this exact one. Note that the dependence on the winding number $n$ appears only in the form of
$|n|$. This fact just means that the internal space is axisymmetric. This mode sum will be the starting point for our analysis of the zeta function.

The coefficients in the one-loop effective potential,
Eq.~(\ref{effpot}), can be written in the form
\begin{eqnarray}
 \int d^2 x A_4(r,\delta_+)
 &=& \int d^2 \tilde x \frac{ A_4(r,\delta_+)}{\rho_+}
 =  -\int_0^1 d\epsilon
       a_4(f=\frac{1}{2}\ln(\frac{2\rho}{\rho_+}))
  -\frac{1}{2}\zeta'(0,\Delta_{4,\epsilon=0})
\,,
      \nonumber\\
\int d^2 x B_4(r,\delta_+)
 &=&\int d^2 \tilde x \frac{B_4(r,\delta_+)}{\rho_+}
  =\frac{1}{2}\zeta(0,\Delta_{4,\epsilon=0})
  =\frac{1}{2} a_4 (f=1)\,,  \label{4dcoef}
\end{eqnarray}
where the zeta function $\zeta(s)$ is defined for the given mass
spectrum in the unwarped frame Eq. (\ref{eigen}):
\begin{eqnarray}
(2\pi)^2\zeta (s)
&= &\int d^2 x  \int d^2 k \sum_{m,n} \lambda_{mnk}^{-2s}\,.
\end{eqnarray}
%%%%%%%%%%%%%%%%%%%%%%%%%%%%%%%%%%%%%%%%%%%%%%%%
The coefficients other than $-\zeta'(0)/2$ in the conformal frame were
 derived in Ref.~\cite{mns}, where it was shown that the term
$A_4(r,\delta_+)$ is given by the equation
%%%%%%%%%%%%%%%%%%%%%%%%%%%%%%%%%%%%%%%%%%%%%%%%
\begin{eqnarray}
  A_4(r,\delta_+)
= \big(A_4(r,\delta_+)\big)_{\rm cocycle}
+ \big(A_4(r,\delta_+)\big)_{\rm unwarped}\,,
\label{a_tot}
\end{eqnarray}
where the first term is the cocycle correction.
In the next section, we will discuss the evaluation of the second term, which is $-\zeta'(0)/2$ in the unwarped, factorizable frame:
\begin{eqnarray}
\int d^2 x \Big( A_4(r,\delta_+)\Big)_{\rm unwarped}
= -\frac{1}{2}\zeta' (0,\Delta_{4,\epsilon=0})\,.
\end{eqnarray}

The cocycle function can be derived from heat kernel
analysis \cite{mns}:\footnote{See also \cite{Hoover},
for UV effects and heat kernel coefficients in higher dimensions.}
\begin{eqnarray}
 A_{4, {\rm cocycle}}(r,\delta_+)
&=&
-\frac{g^2(2\pi-\delta_+)}{1440\pi^2}
\int_0^1 d\epsilon
\int^{\infty}_0 dz
\Big(\frac{1}{2}\ln \Big(\frac{2\rho}{\rho_+}\Big) \Big)
\frac{z}
     {(1+z^2)(1+r z^2)}
\frac{\Psi(1,r,\epsilon, z)}{(1+r z^2)^4}
\nonumber\\
&-&\frac{g^2}{144\pi}
\frac{\delta_+}{2\pi}
\frac{2-\frac{\delta_+}{2\pi}}
     {1-\frac{\delta_+}{2\pi}}
 \Big[2 \ln(2)
       -1+r
      -\frac{1}{5}
       \frac{1+(1-\frac{\delta_+}{2\pi})^2}
            {(1-\frac{\delta_+}{2\pi})^2}
        \Big(
         \frac{1}{2}\ln (2)  (1-3r)
          +\frac{1-r}{2}
        \Big)
 \Big]
 \nonumber\\
&-& \frac{g^2}{144\pi}
\frac{\frac{\delta_+}{2\pi}-(1-r)}{r}
\frac{r+(1-\frac{\delta_+}{2\pi})}
     {1-\frac{\delta_+}{2\pi}}
\nonumber\\
&\times&
 \Big[2\ln(2r)
       -1+\frac{1}{r}
      -\frac{1}{5}
       \frac{r^2+(1-\frac{\delta_+}{2\pi})^2}
            {(1-\frac{\delta_+}{2\pi})^2}
          \Big(
           \frac{1}{2}\ln(2r) (1-\frac{3}{r})
            +\frac{1-\frac{1}{r}}{2}
          \Big)
 \Big] \,,  \label{inflationary2}
\end{eqnarray}
where $\Psi(\rho_+,\rho_-,\epsilon,z)$ can be found in Appendix B of \cite{mns}
and is just a function of $r$ and $z$. The independence of $\Psi$
on $\epsilon$
is discussed in \cite{mns}. The coefficient $B_4(r,\delta_+)$ is given by
\begin{eqnarray}
B_4(r,\delta_+)=\Big(B_4(r,\delta_+)\Big)_{\rm bulk}
              +\Big(B_4(r,\delta_+)\Big)_{\rm brane}\,,
\label{bcoef}
\end{eqnarray}
where
\begin{eqnarray}
&&\big(B_4(r,\delta_+)\big)_{\rm bulk}
=
\frac{g^2(2\pi-\delta_+)}{1440\pi^2}
\int^{\infty}_0 dz
\frac{z}
     {(1+z^2)(1+r z^2)}
\frac{\Psi(1,r,0, z)}{(1+r z^2)^4}
\,, \label{bulk}
\end{eqnarray}
and
\begin{eqnarray}
 \big(B_4(r,\delta_+)\big)_{\rm branes}
&=& \frac{g^2}{288\pi}
\frac{\delta_+}{2\pi}
\frac{2-\frac{\delta_+}{2\pi}}
     {1-\frac{\delta_+}{2\pi}}
 \Big[4
      -\frac{1}{5}
       \frac{1+(1-\frac{\delta_+}{2\pi})^2}
            {(1-\frac{\delta_+}{2\pi})^2}
          (1-3r)
 \Big]
 \nonumber\\
&&
+  \frac{g^2}{288\pi}
\frac{\frac{\delta_+}{2\pi}-(1-r)}{r}
\frac{r+(1-\frac{\delta_+}{2\pi})}
     {1-\frac{\delta_+}{2\pi}}
 \Big[4
      -\frac{1}{5}
       \frac{r^2+(1-\frac{\delta_+}{2\pi})^2}
            {(1-\frac{\delta_+}{2\pi})^2}
          (1-\frac{3}{r})
 \Big]\,,
\label{inflationary}
\end{eqnarray}
which is also derived from heat kernel analysis.

\section{Zeta function regularization in the unwarped conformal frame}

Given the eigenvalues we found in the previous section, we shall now derive
an expression for the corresponding spectral zeta function and related quantities, such as the effective action. 
In Ref \cite{mns} the authors used the density of states method because they had not found an exact mode spectrum.

The first step on our journey will be to use the Mellin transform:
\begin{eqnarray}
\zeta (s)={1\over \Gamma(s)} \int d^2 x  \int d^2 k \sum_{m,n} \int_0^\infty t^{s-1}
\exp\{-[k^2 +g^2( a(m+\alpha)^2 +b(m+\alpha)|n| + c n^2 +q)]t\}\,,
\label{zetafunk}
\end{eqnarray}
where we have just rearranged the eigenvalue $\lambda_{mnk}$, see
equation (\ref{eigen}), such that it takes the form of a two
dimensional Epstein zeta function, and
\beq
a=4, \qquad\qquad b= {4(1+r)\over \kappa},\qquad\qquad c={4r\over \kappa^2}, \qquad\qquad
q=-1,\qquad\qquad \alpha=1/2\,.
\eeq
Firstly, we perform the $k$-integration after interchanging the order of integrations:
\begin{eqnarray}
\int d^2 x  \int d^2 k \sum_{m,n} \lambda_{mnk}^{-2s}
&=&{1\over \Gamma(s) } \int d^2 x  \int d^2 k \sum_{m,n}
\int_0^\infty dt t^{s-1}
\exp\{-[k^2 +g^2( a(m+\alpha)^2 +b(m+\alpha)|n| + c n^2 +q)]t\}
\nonumber\\
&=&
\int d^2 x \sum_{m,n} \frac{2 \pi}{\Gamma(s)} \int_0^{\infty}
dt t^{s-2} \exp\{-g^2 [a(m+\alpha)^2 +b(m+\alpha)|n| + c n^2 +q]t\}
\nonumber\\
&=&\int d^2x
\frac{2\pi g^{2(1-s)}}{s-1}\sum_{m,n}
 [a(m+\alpha)^2 +b(m+\alpha)|n| + c n^2 +q]^{1-s}
 \label{modesum}
\end{eqnarray}
where in the third step we have redefined the terms such that
\begin{eqnarray}
&&\hat \alpha(n) =\alpha+\frac{b n}{2a}
              =\frac{1}{2}+\frac{(1+r)|n|}{2\kappa},
\nonumber\\
&&\hat q(n) =q + c n^2-\frac{(bn)^2}{4a}
         = -\frac{n^2}{\kappa^2} (1-r)^2 -1
\,. \label{nsum}
\end{eqnarray}
This form shows that the zeta function is essentially some kind of
two-dimensional Epstein zeta function as we show below. In the following discussion, we focus on the non-axisymmetric modes
$n\neq 0$. For the contribution of the axisymmetric modes, see Appendix A.

\subsection{Extended binomial expansion}

Frequently, the form of the two-dimensional Epstein zeta function
allows one to perform its summation in an elegant way which involves
the Chowla-Selberg expansion formula or more frequently (as it would
correspond to the case here) a generalization thereof, see
\cite{ElizaldeCS}. However, some conditions must be fulfilled, in
particular, the quadratic form must be positive definite and the
constant $q$ term should be also non-negative. But this is not here
the case, and the fact that
  $q<0$ does not allow for such a beautiful analysis.
 Henceforth, we shall apply in what follows what we have called the
{\it extended binomial expansion} approach (see also \cite{CM2}), namely, we
will introduce an extra summation via
\begin{eqnarray}
&& \sum_{m=0}^{\infty}
   \sum_{n=1}^{\infty}
 \big[a(m+\alpha)^2+b(m+\alpha)n+c n^2+q \big]^{-s+1}
 \nonumber\\
&=&\sum_{m=0}^{\infty}
   \sum_{n=1}^{\infty}
 \big[a(m+\hat \alpha)^2+\hat q \big]^{-s+1}
=\sum_{m=0}^{\infty}
   \sum_{n=1}^{\infty}
   \sum_{j=0}^{\infty}
  \frac{\Gamma(j+s-1)}
       {\Gamma(s-1)j!}
  \big[a(m+\hat \alpha)^2\big]^{1-s-j}
  {\hat q}^{j}\,,
\end{eqnarray}
where the validity of the binomial expansion is defined for
\begin{eqnarray}
  \Big|
\frac{\hat q}{a(m+\alpha)^2}
  \Big|<1\,,
\end{eqnarray}
which is indeed satisfied for all possible values of $m$ and $n$ in
our model.

The crucial step is to interchange now the summation over $j$ with
 the $m,n$ summations. The absolute convergence of the
binomial expansion in $j$ guarantees such an
interchange. Whence, we obtain that
\begin{eqnarray}
\sum_{m=0}^{\infty}
   \sum_{n=1}^{\infty}
 \big[a(m+\alpha)^2+b(m+\alpha)n+c n^2+q \big]^{-s+1}&=&
\sum_{j=0}^{\infty}\sum_{n=1}^{\infty}
 a^{1-s-j}
\frac{\Gamma(j+s-1)}
      {\Gamma(s-1)j!}
      {\hat q}^{j}
      \zeta_{H}(2s+2j-2,\hat \alpha)\,,
\end{eqnarray}
where we have summed over $n$ to give a standard relation in terms of
Hurwitz zeta functions.  Substitution of the above relation into
Eq.~(\ref{modesum}) leads to
\begin{eqnarray}
\zeta(s)\big|_{n\neq 0}
&=&{g^{2(1-s)}\over \pi} \int d^2x
 \sum_{j=0}^{\infty} G(j,s) \sum_{n=1}^{\infty}
      \Big(\frac{n^2}{\kappa^2} (1-r)^2 +1\Big)^{j}
      \zeta_H\Big(2s+2j-2,\frac{1}{2}+\frac{1+r}{2\kappa}n\Big)
      \label{zetaS}
      \nn
\end{eqnarray}
with
\beq
G(j,s)=\frac{ 2^{-2(j+s-1)} \Gamma(s+j-1)}
      {\Gamma(s)j!}\,.
\label{Gee}
\eeq
Note also that the presence of $|n|$ in our mode sum, with
$-\infty \leq n\leq \infty$, makes the zeta function a non-typical
one. In our approach, we have just multiplied by two and
taken the sum from $n=1\dots\infty$, the $n=0$ mode having been
treated independently.

The above expression does not converge unless Re $s\geq 3$ and thus,
we must first analytically continue the zeta function in equation
(\ref{zetaS}) in the usual way \cite{ElizaldeBook}, by first
subtracting off the terms responsible for the divergences in
$\zeta(s)$. After this we are able to take the derivative at $s=0$,
being careful to add  back the terms we subtracted off
before the analytic continuation.

\subsection{Analytic continuation of the $\zeta\,-$~function}

In order to isolate the divergent behavior in the zeta function,
due to small $s$, we need to make an asymptotic expansion of
Eq.~(\ref{zetaS}),
the details of which are given in Appendix A.
From this we can
 now analytically continue $\zeta(s)$ and take the derivative at $s=0$.
Subtracting off the divergent terms from the zeta function leads to a
function
\bea
P(s)
&=&
{g^{2(1-s)}\over \pi} \int d^2x\sum_{j=0}^\infty G(j,s) \sum_{n=1}^{\infty}
\Big[\Big(\frac{n^2}{\kappa^2} (1-r)^2 +1\Big)^{j}
      \zeta_H\Big(2s+2j-2,\frac{1}{2}+\frac{1+r}{2\kappa}n\Big) -F(n,j;s)
\Big],
\eea
where $P(s)$ is now a regular function at $s=0$ and $F(n,j;s)$ is defined in Appendix
A. Performing
the sum over $j$ and adding back the subtracted terms leads
to the analytically continued zeta function
\beq
\zeta(s)\Big|_{n\neq 0}= P(s) + \sum_{j=0}^\infty G(j,s) \Delta(j,s)
\eeq
where, given that the subtracted terms are simple functions of $n$,
they can be expressed as a combination of ordinary Riemann zeta
functions:
\beq
\Delta(j,s) = {g^{2(1-s)}\over \pi} \int d^2 x\,
{\kappa^{1+2s}}  {(1-r)^{2j}\over
 (1+r)^{1+2j+2s} }
\frac{2^{-6+2j+2s}}{45(2s+2j-3)}
 \Big[ {w_0\over\kappa^4} \zeta_R(2s-3) +{w_2(j,s)\over\kappa^2} \zeta_R(2s-1)
+w_4(j,s) \zeta_R(2s+1)\Big] \label{Djs},
\eeq
the functions $w_0,w_2$ and $w_4$ being defined in Appendix A. Note,
however, that we must leave the sum over $n$ in $P(s)$ as it is. The
$j=0$ and $j=1$ terms must be considered separately because of the
gamma functions in $G(j,s)$ (see Eq.~(\ref{Gee})), which leads to
compensations of poles and zeros of the Riemann zeta functions in
Eq.~(\ref{Djs}). Taking all of this into account, the analytic
continuation of the derivative of $P(s)$ with respect to $s$ can be
 found without any further problems.

Then, after analytic continuation to $s\to 0$, we obtain $P'(0)$ as
given in Appendix B.
In Fig.~1 we have plotted the function $P'(0)$ as a function of $r$.
As can be seen, the summation over $j$ converges fairly rapidly
and the result for $j_{\rm max}=10$ already coincides
well with that for $j_{\rm max}=50$.
In the following calculations, we shall use $j_{\rm max}=50$ as a
conservative choice. Note that in all of our plots the $n$-summation
 converges fairly quickly due to the asymptotic expansion and
the first few terms would suffice;
however, we will always set $n_{\rm max}=50$ as a conservative choice.

%%%%%%%%%%%%%%%%%%%%%
\begin{figure}
\begin{center}
  \begin{minipage}[t]{.45\textwidth}
   \begin{center}
    \includegraphics[scale=.75]{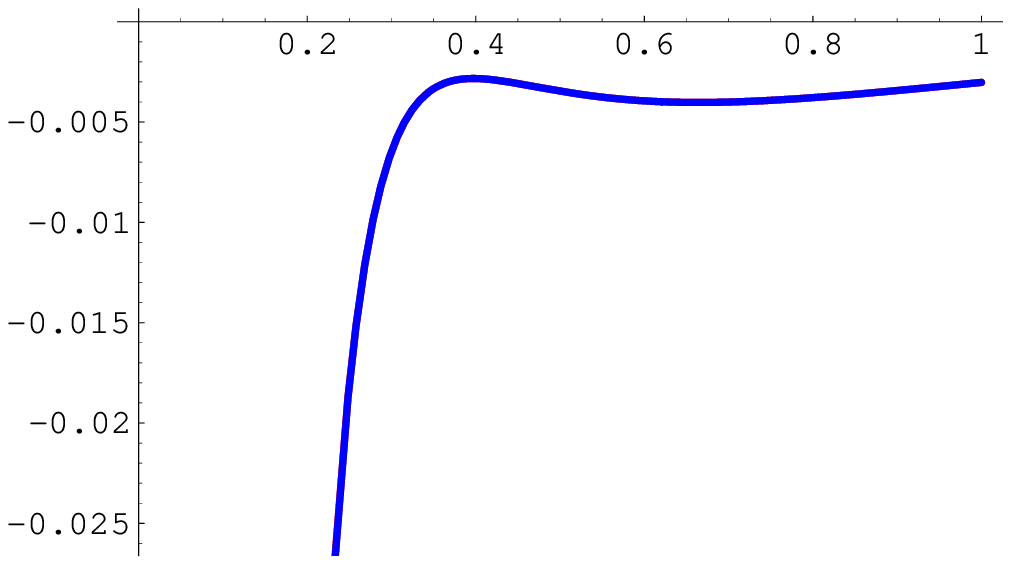}
        \caption{Plots for $ P'(0)$, Eq.~(\ref{Pprime}), as a function
of $r$ are shown for $\delta=0.01$. The red and blue curves correspond
to the truncation of the $j$-summation at $j_{\rm max}=10$ and
$j_{\rm max}=50$, respectively.
       }
   \end{center}
   \end{minipage}
\hspace{0.5cm}
   \begin{minipage}[t]{.45\textwidth}
   \begin{center}
    \includegraphics[scale=.75]{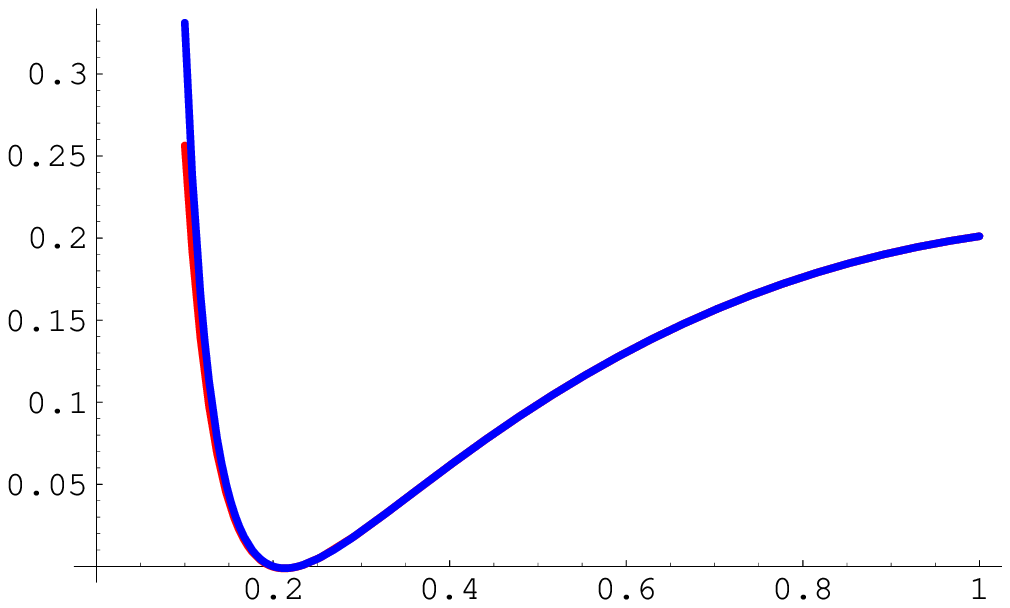}
\caption{Plots of the derivative
$(d/ds) (\sum_j G\Delta)(s)|_{s\to 0}$, Eq.~(\ref{GDprime}), as a
function of $r$ are shown for $\delta=0.01$.
The red and blue curves correspond to the truncation of the
$j$-summation at $j_{\rm max}=10$ and $j_{\rm max}=50$,
respectively.}
   \end{center}
   \end{minipage}
   \end{center}
\end{figure}
%%%%%%%%%%%%%%%%%%%
The analytic continuation of the counterterms $\sum_j G(j,s)\Delta(j,s)$ is discussed in Appendix B, where after considering the $j=0$, $j=1$ and $j\geq 2$ modes separately we find (see App.~B)
\begin{eqnarray}
\frac{d}{ds}\sum_{j=0}^{\infty}\Big(G(j,s)\Delta(j,s)
\Big)\Big|_{s= 0}
=
 \frac{d}{ds}\Big(G(0,s)\Delta(0,s)\Big)\Big|_{s= 0}
+\frac{d}{ds}\Big(G(1,s)\Delta(1,s)\Big)\Big|_{s= 0}
+\sum_{j=2}^{\infty}\frac{d}{ds}\Big(G(j,s)\Delta(j,s)\Big)
\Big|_{s= 0}
\label{GDprime}
\end{eqnarray}
In Fig.~2 we plot Eq.~(\ref{GDprime}) as a function of $r$. The
$j$-summation truncated at $j_{\rm max}=10$ already exhibits
good convergence and coincides well with that for $j_{\rm max}=50$.
In the following calculations, we use $j_{\rm max}=50$ as a
conservative choice.

We finish this section by noting that the above, quite involved
mathematics have yielded in the end very precise numerical results.
We will now take advantage of them by considering their implication
---e.g., those of the Casimir corrections from bulk fields--- for the
phenomenological predictions that were mentioned in the first
section.

\section{Implications for the hierarchy problem}

We can now consider the physical implications
on the brane in the original 6D theory given by Eq. (\ref{theory6d}).
In Ref \cite{mns},
several phenomenological applications have been discussed, i.e.,
to the hierarchy problem and the vacuum (Casimir) energy density
(i.e., the effective 4D cosmological constant) on the branes.
With our new results at hand, we give a revised version of them,
 especially focusing on the hierarchy problem.
After analysing the hierarchy problem,
we briefly discuss the case of the vacuum energy density.

In the original six-dimensional model, the {\it effective}
four-dimensional Planck scale is
 \begin{eqnarray}
   M_{\rm pl}^2
 \simeq
 \frac{\rho_+(2\pi-\delta_+)}{g_0^2}
  M_{6}^4\,.
 \end{eqnarray}
 If we assume a brane localized field whose bare mass is given by
$m^2$  on either brane at $\rho_{\pm}$ then the observed mass scales
are $ m_{+}^2= m^2$ and $m_{-}^2= r^2 m^2\,$.
Thus, the mass ratio between the field and the effective Planck mass
is given by\begin{eqnarray}
\frac{m_{\pm}^2}{M_{\rm pl}^2}
\simeq
 \frac{\mu^2 m^2}{M_{6}^4}\
  \frac{g_0^2}{2\pi-\delta_+}
   e^{-(2A_6 +  B_6)/(2B_6)}\,,
\end{eqnarray}
where $A_6$ and $B_6$ correspond to $A_4$ and $B_4$ in the 6D model,
respectively.
Assuming the factor $(\mu m/M_{6}^2)^2$ takes an optimal value of
 ${\cal O}(1)$ at the unification of the fundamental scales, then
the mass ratio becomes
\begin{eqnarray}
  \frac{g_0^2}{2\pi-\delta_+}
   e^{-(2A_6 + B_6)/(2B_6)}\,,
\end{eqnarray}
where we have used the value of $\rho_+$, at stabilization.
The corresponding quantity in the 4D toy model is given by
\begin{eqnarray}
R(r,\delta_+)
 = \frac{g^2}{2\pi-\delta_+}
   e^{-(A_4+B_4)/B_4}\,.\label{hier}
\end{eqnarray}
%%%%%%%%%%%%%%%%%%%%%%%%%%%%%
\begin{figure}
\begin{center}
  \begin{minipage}[t]{.45\textwidth}
   \begin{center} \vspace*{-1mm}
    \includegraphics[scale=.85]{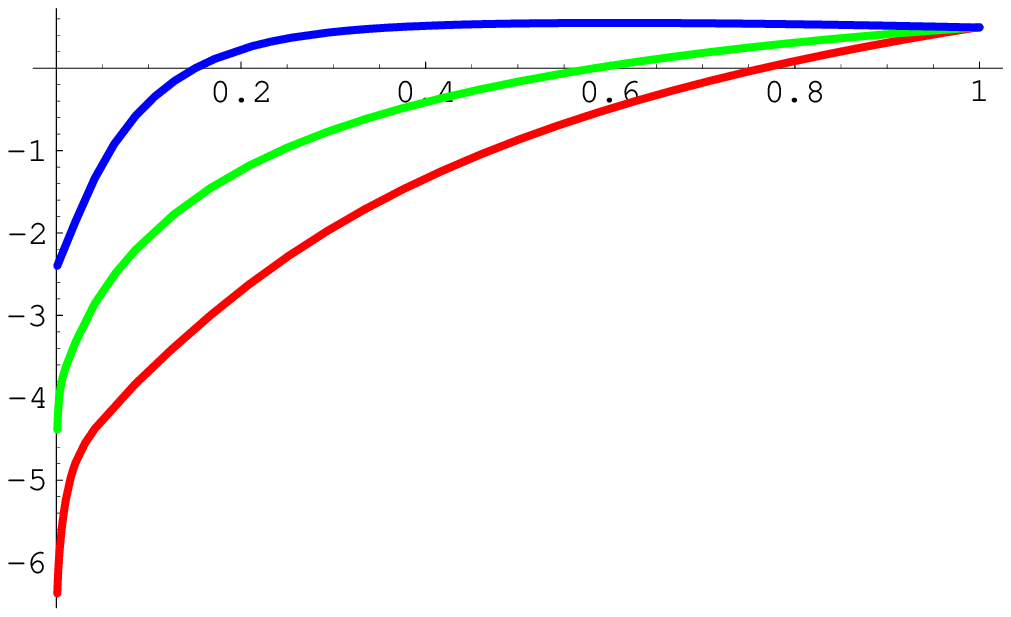}
        \caption{Numerical plots of $ \log_{10}R(r,\delta_+=0.01)$ as
a function of $r$ are shown for $g=0.5,5,50$ (the red, green and blue
curves, respectively). We set $j_{\rm max}=n_{\rm max}=50$ as a
conservative choice.}
   \end{center}
   \end{minipage}
   \hspace{9mm}
   \begin{minipage}[t]{.45\textwidth}
   \begin{center}
    \includegraphics[scale=.83]{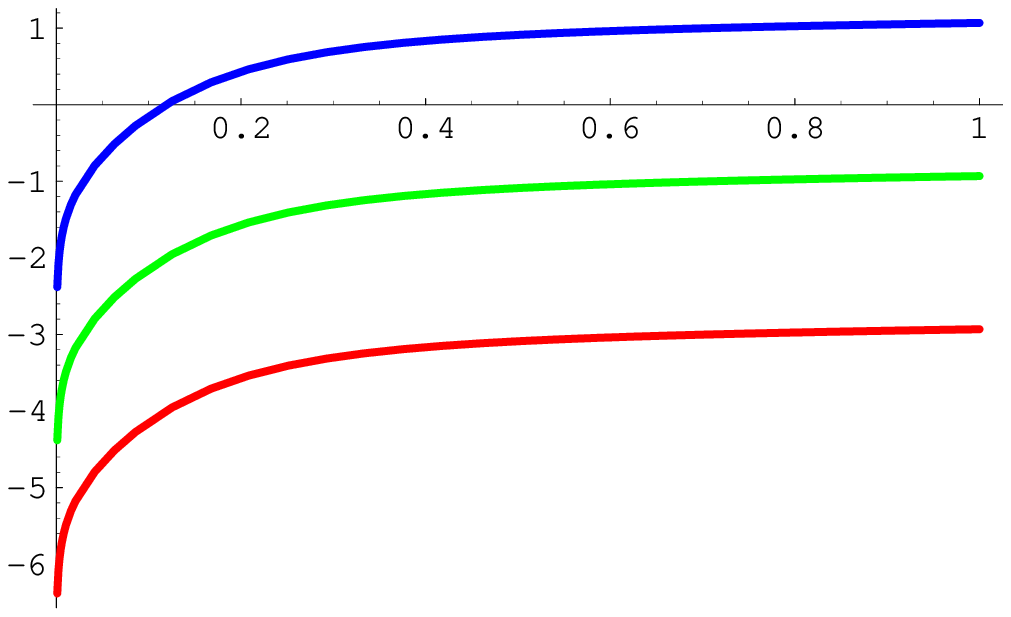}
\caption{Numerical plots of $ \log_{10}R(r,\delta_+=0.01)$, via the
WKB approximation, as a function of $r$ are shown for $g=0.5,5,50$
(red, green and blue curves, respectively).}
   \end{center}
   \end{minipage}
   \end{center}
\end{figure}
%%%%%%%%%%%%%%%%%%%%%%

In Fig.~3 we have plotted $\log_{10}\big[R(r,0,01)\big]$ as a
function of $r$. The resulting ratios between the scalar and
gravitational energies lead to the same conclusions as in the
analysis given in Ref. \cite{mns}, in which the derivative of the zeta function was
evaluated by the WKB method (See Fig. 4). Namely, we obtain larger mass
hierarchies for smaller $r$ and smaller $g$; in detail, however, the
form of $R$ is not the same. The most important difference is the
behavior at $r=1$: in the exact spectrum $R$ does not depend on the
coupling $g$ (whereas in the WKB approximation it scales as $g^2$).
Numerical plots also show that this feature is almost independent of
the value of the deficit angle $\delta_+$. Thus, the expected mass
hierarchy in the rugby-ball frame is independent of the value of the
dilaton coupling $g$, which is actually the bulk cosmological
constant, because the dilaton dynamics are absent in the conformal
(rugby-ball) frame.

Finally, we add a few comments on the Casimir energy density
realized on the branes, i.e., the effective 4D cosmological
constant. As shown in Eq. (56) (also see (57)), the expression for
the vacuum energy density has a very similar form to $R$, since
$B_4(r,\delta_+) \propto g^2$. In fact, after setting $\mu= M_4$ as
an optimal choice,
 numerical plots of Eq. (57) in Ref. \cite{mns} as a function of $r$
for different dilaton coupling $g$ again show convergence to the
same point in the $r\to 1$ limit, though here we do not present
these results explicitly. For smaller $r$, the same behavior as in
Fig.~5 in \cite{mns} is obtained and implies a tiny amount of
effective cosmological constant on the branes, in comparison with
the gravitational energy scale $M_4^2$. The only problem is that the
sign of the energy density becomes negative, as is expected from the
form of the effective potential Eq. (\ref{effpot}). This fact
requires an additional uplifting mechanism for the effective
potential.

%%%%%%%%%%%%%%%%%%%%%%%%%%%%
\section{Conclusion}

In this article, we have given a precise analysis of the mode
summation in deriving the one-loop effective potential for the
volume modulus in a 4D analogue model of a warped flux
compactification based on 6D supergravity. In \cite{mns} the mass
spectrum was found using the leading order WKB approximation.
Strictly speaking, this is only an approximation and an exact
one-loop analysis of the mass spectrum has been performed in the
present paper, by employing zeta function regularization techniques.
We would also like to stress that the unwarped frame is not the
standard {Kaluza-Klein} spacetime, because of the presence of
conical singularities.\footnote{A beautiful discussion of zeta
function regularization in standard KK theories can be found in
\cite{CM2}.}

In this paper we used our results to look at the hierarchy problem in
warped flux compactification models. The qualitative features
in Fig.~3 are similar to those of Fig.~4: we expect a larger mass
hierarchy for smaller $r$ (where $r$ is the degree of warping) and
for larger bulk dilaton coupling $g$. Indeed, for $r\ll 1$, the
contribution from the cocycle function becomes important and gives
rise to a large mass hierarchy on the brane, though at the same time
the effective mass of the modulus may become lighter (and hence more
unstable to KK perturbations). However, quantitatively there is a
significant difference for $r\lesssim 1$, where quite surprisingly all
the curves for different values of $g$ in Fig.~3 converge to the same
point at $r=1$. In Fig.~4, which is for the WKB approximation, the
behavior of $R$ is proportional to $g^2$, even at $r=1$. This may well
represent a non-trivial feature of the Casimir effect in the rugby
ball limit.
Similar results are obtained in the case of the vacuum (Casimir)
energy density on the branes, namely convergence to the same value
in the $r\to 1$ limit, independently of $g$. For smaller values of
$r$, we can expect a tiny amount of the effective cosmological
constant on the branes, in comparison with the gravitational energy
scale. The negativity of the energy density may require slight
modifications of the background model.

As a consistency check of our results, in Figs.~5 and 6 we have
plotted the integrands of  $\zeta(0)$ and $B_4(r,\delta_+)$, which
are related to $a_4(f=1)$ given in Eq.~(\ref{4dcoef}), as functions
of $g$ and $r$, respectively, for a fixed value of the deficit
angle, $\delta_+$. Note that both quantities are practically
insensitive to $\delta_+$. In the unwarped frame, due to the
properties of the heat kernel coefficients (see e.g., \cite{VSV}),
the equation $\zeta(0)= a_4(f=1)$ should be satisfied. Although they
do not agree exactly, they do exhibit a similar behavior, namely
increasing for larger $g$ and for smaller $r$. We thus conclude that
our results are physically reliable. Nevertheless, there are still
slight differences between these values and it is worthwhile to
consider the possible origin of this difference.
%%%%%%%%%%%%%%%%%%%%%%%%%%%%%
\begin{figure}
\begin{center}
  \begin{minipage}[t]{.45\textwidth}
   \begin{center}
    \includegraphics[scale=.65]{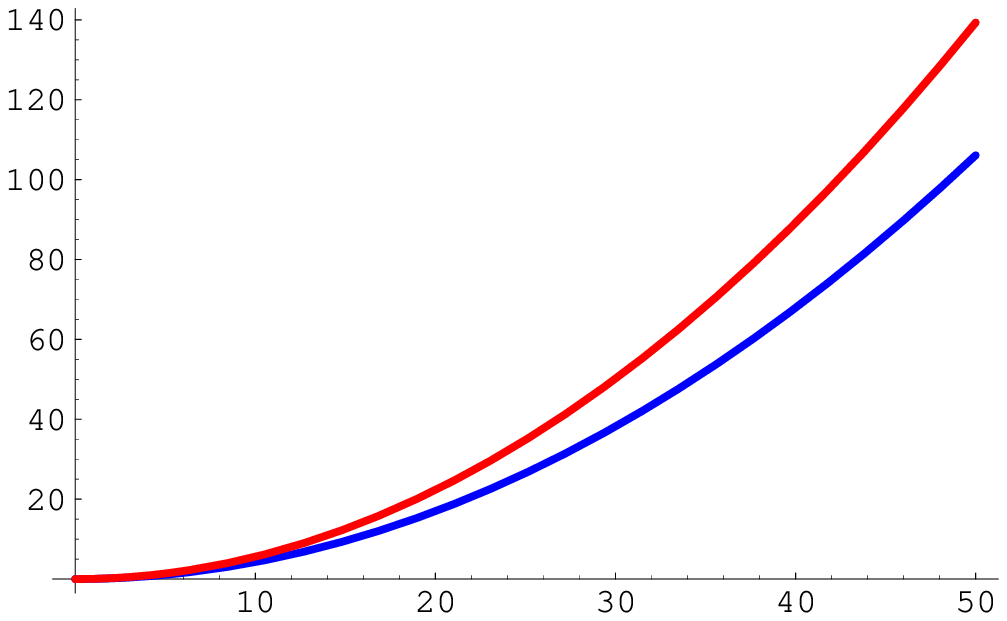}
        \caption{Numerical plot of the integrand of $\zeta(0)$ and
$2 B_4$ as functions of $g$ are shown for $r=1$ and $\delta_+=0.01$.
The red and blue curves correspond to the cases of $\zeta(0)$ and
$2B_4$, respectively. We set $j_{\rm max}=n_{\rm max}=50$.}
   \end{center}
   \end{minipage}
   \hspace{0.5cm}
   \begin{minipage}[t]{.45\textwidth}
   \begin{center}
    \includegraphics[scale=.65]{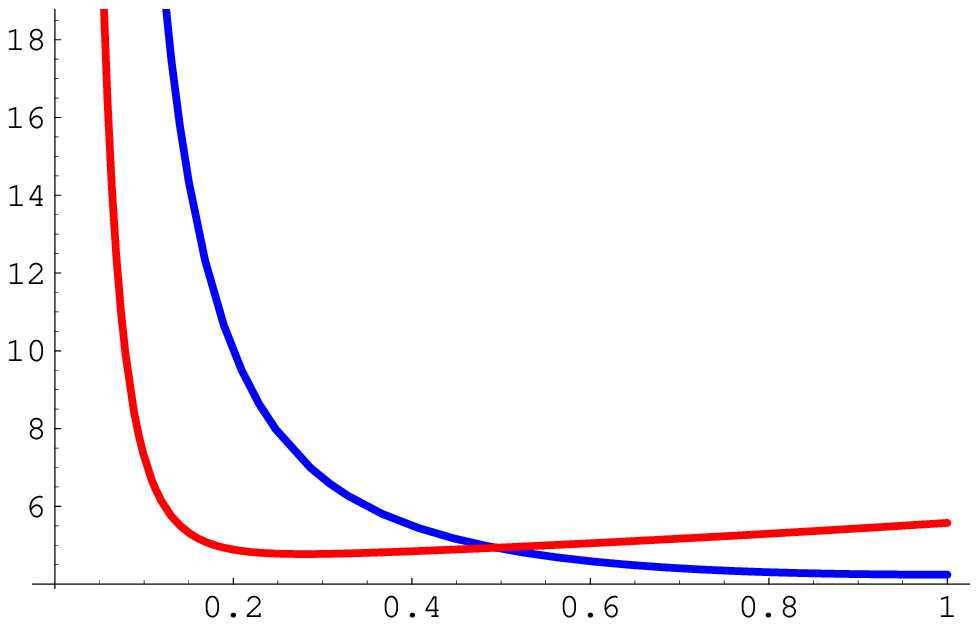}
\caption{Numerical plot of the integrand of $\zeta(0)$ and $2 B_4$
as functions of $r$ are shown for $g=10$ and $\delta_+=0.01$. The red
and blue curves correspond to the cases of $\zeta(0)$ and $2B_4$,
respectively. We set $j_{\rm max}=n_{\rm max}=50$.}
   \end{center}
   \end{minipage}
   \end{center}
\end{figure}
%%%%%%%%%%%%%%%%%%%%%%

%%%%%%%%%%%%%%%%%%%%%%%%corrections we -> the authors %%%%%%%%%%%%%%%%%%%%%%%%
In \cite{mns} the authors confirmed that $B_4$ (and thus $a_4(f=1)$) is
independent of the value of $\epsilon$ \cite{VSV}, which
characterizes the conformal transformation given in Eq.
(\ref{confort4d}) and therefore, we believe that $B_4$ is
correct.\footnote{If any terms in $B_4$ were incorrect then
conformal invariance would be ruined.} 
%%%%%%%%%%%%%%%%%%%%%%%%%%
We have also carefully
checked the convergence of the mode summations. Thus, it is the view
of the authors that a possible origin for the disagreement could be
in our determination of the exact mode spectrum. By imposing
vanishing conditions for the mode functions and their derivatives at
the poles (branes), we were able to derive the mass spectrum in the
rugby-ball frame. Our method to determine the mass spectrum is
essentially based on the same arguments given in
\cite{Parameswaran:2006db}, where two conditions, normalizability
and regularity at the poles were imposed (more precisely, it was
demanded that the wave functions should be Hermitian). However, if
only normalizability were imposed, this would allow for logarithmic
divergences at the poles and this may well lead to additional modes
in the eigenvalue spectrum, which could then reflect in the
differences between Figs.~5 and 6. Regardless of this, based on the
qualitative behavior of $B_4$ and $\zeta(0)$, the existence of such
modes does not seem likely to significantly affect the qualitative
behavior of our results.

In closing we would also like to stress that although the results
presented here are strictly for a toy 4D warped flux
compactification model, the methods introduced can be generalized
straightforwardly to six dimensions. However, as discussed in
\cite{mns}, to start with the relevant heat kernel coefficient on
the cone in six dimensions should be found. In addition, in the
current investigation we have focused on the simplest case: a
massless, minimally coupled scalar field. It should be also possible
to extend our analysis to the case of a bulk scalar field with self
interactions and other fields in the multiplets appearing in the
supergravity model, which our model is based on. We hope to report
on the results of these extended analyses in the near future.

%%%%%%%%%%%%%%%%%%%%%%%%%%%%%%%%%%%%%%%
\section*{Acknowledgements}
EE was supported in part by MEC (Spain), projects PR2006-0145 and
FIS2006-02842, and by AGAUR, contract 2005SGR-00790. MM was supported
in part by Monbu-Kagakusho Grant-in-Aid for Scientific Research (B)
No. 17340075 at the YITP and by the project ``Transregio (Dark
Universe)" at the ASC.

%%%%%%%%%%%%%%%%%%%%%%%%%%%%%%%%%%%%
\appendix

\section{Asymptotic expansion of the zeta function}

In order to isolate the divergent behavior in the zeta function,
due to small $s$, we need to expand Eq.~(\ref{zetaS}).
Towards this end we shall make use of the asymptotic
expansion of the Hurwitz zeta function \cite{ElizaldeBook}:
\begin{eqnarray}
\zeta_{H}(2s+2j-2,\frac{1}{2}+\frac{1+r}{2\kappa}n)
&=&\frac{1}{2s+2j-3}
 \Big(\frac{1}{2}+\frac{(1+r)n}{2\kappa}\Big)^{-(2s+2j-3)}
+\frac{1}{2}
     \Big(
     \frac{1}{2}+\frac{(1+r)n}{2\kappa}
     \Big)^{-(2s+2j-3)-1}
\nn
&+&\frac{1}{2s+2j-3}
\sum_{k=2}^{\infty}
   \frac{B_k}{k!}(2s+2j-3)_k
   \left(\frac{1}{2}+\frac{1+r}{2\kappa}n\right)^{-(2s+2j-3)-k}\,,
\end{eqnarray}
where
the $B_k$ are Bernoulli numbers. Furthermore, in the limit of large
$n$
and  fixed $j$, the asymptotic expansion of $\zeta_{H}$  is found to be
%%%%%%%%%%%%%%%%%%%%%%%%%%%%%%%%%%%%%%%%%%
\beq
\zeta_{H}(2s+2j-2,\frac{1}{2}+\frac{1+r}{2\kappa}n)
\Big[
 \frac{n^2}{\kappa^2}(1-r)^2
+1
\Big]^j\simeq F(n,j;s),
\eeq
where
\beq
F(n,j;s) :=
 \frac{n^{-1-2s}}{\kappa^{-1-2s}}
{(1-r)^{2j}\over
(1+r)^{1+2j+2s}}
\frac{2^{-6+2j+2s}}{45(2s+2j-3)}
\
\Big( w_0 \frac{ n^4}{\kappa^4}
     +w_2 \frac{n^2}{\kappa^2}
     +w_4 \Big)
\,,
\eeq
and we have defined
\begin{eqnarray}
w_0(j,s)&=& 360 (1+r)^{4}\,,
\nonumber\\
w_2(j,s)&=&\frac{360 j (1+r)^4}{(1-r)^2}
    -60 (1+r)^2(-3+2s+2j)(-2+2s+2j)\,,
\nonumber\\
w_4(j,s)&=& \frac{180(j-1)j(1+r)^4}{(1-r)^4}
-\frac{60j(1+r)^2(-3+2j+2s)(-2+2s+2j)}{(1-r)^2}
\nonumber\\
&+&7 (-3+2s+2j)(-2+2s+2j)(-1+2s+2j)(2s+2j)\,.
\end{eqnarray}
Note the form of these terms, especially $w_4$, will be important
for the analytic continuation of the zeta-function.

%%%%%%%%%%%%%%%%%%%%%%%%%%%%%%

\section{Analytic continuation}

In this appendix we discuss the analytic continuation of the various
functions discussed in the main text.
The expression for the analytic continuation of $\zeta(s)$ to $s\to 0$
is given by
\begin{eqnarray}
\zeta(0)
&=&
 \frac{g^2}{\pi}
\int d^2x
\sum_{n=1}^{\infty}
\Big[
-4\Big(\zeta_{H}(-2,\frac{1}{2}+\frac{1+r}{2\kappa}n)
 -F(n,0,0)
\Big)
+\Big(\zeta_{H}(0,\frac{1}{2}+\frac{1+r}{2\kappa}n)
   \big(\frac{n^2}{\kappa^2}(1-r)^2+1\big)
 -F(n,1,0)
\Big)
\Big]
\nonumber \\
&+&
\frac{\kappa g^2}{720\pi(1+r)}
\int d^2x
\Big[
\Big(
 \frac{(1+r)^4}{\kappa^4}
-\frac{(1+r)^2}{\kappa^2}
-14
\Big)
-\frac{(1-r)^2}{(1+r)^2}
\Big(
\frac{3(1+r)^4}{\kappa^4}
+\frac{3(1+r)^4}{(1+r)^2}\kappa^2
 -14
\Big)
\nonumber\\
&+&\sum_{j=2}^{\infty}
\frac{(1-r)^{2j}}{(2j-3)(1+r)^{2j}}
\Big(
90\frac{(1+r)^4}{(1-r)^4}
-60\frac{(1+r)^2}{(1-r)^2}(2j-3)
+14(2j-3)(2j-1)
\Big)
 \Big]\,.
\end{eqnarray}
Note that the axisymmetric modes do not contribute to $\zeta(0)$.

We next discuss the analytic continuation that
is required to derive the expression for $\zeta'(0)$.
The analytic continuation of $P'(s)$ to $s\to 0$ is given by
\begin{eqnarray}
P'(0)&=&
\frac{(2g)^{2}}{\pi}
\Big\{
\Big(2\ln(2g)-1\Big)
 \sum_{n=1}^{\infty}
  \Big[\zeta_H(-2,\frac{1}{2}+\frac{1+r}{2\kappa}n)
   -F[n,0;0]
  \Big]
\nonumber\\
&-&\frac{\ln(2g)}{2}
     \sum_{n=1}^{\infty}
     \Big[\zeta_H(0,\frac{1}{2}+\frac{1+r}{2\kappa}n)
     \Big(\frac{n^2}{\kappa^2} (1-r)^2 +1\Big)
   -F[n,1;0]
  \Big]
\nonumber \\
&-& \sum_{n=1}^{\infty}
  \Big[2\zeta_H{}'(-2,\frac{1}{2}+\frac{1+r}{2\kappa}n)
   -\partial_s F[n,0;s]\Big|_{s \to 0}
  \Big]
\nonumber\\
&+&\frac{1}{4}
\sum_{n=1}^{\infty}
\Big[2\zeta_H{}'(0,\frac{1}{2}+\frac{1+r}{2\kappa}n)
    \Big(\frac{n^2}{\kappa^2}(1-r)^2+1\Big)
   -\partial_s F[n,1;s]\Big|_{s\to 0}
\Big]
\nonumber\\
&+&\sum_{j=2}^{\infty}
 \frac{2^{-2j}}{j(j-1)}
\sum_{n=1}^{\infty}
\Big(
 \zeta_H(2j-2,\frac{1}{2}+\frac{1+r}{2\kappa}n)
    \Big(\frac{n^2}{\kappa^2}(1-r)^2+1\Big)^j
   - F[n,j;0]
\Big)
 \Big\}\,.  \label{Pprime}
\end{eqnarray}

Note, we must also pay attention to the counterterms.
The product $G(j,s)\Delta(j,s)$ can be written as
\begin{eqnarray}
G(j,s)\Delta(j,s)
&=&
\frac{1}{2^4\pi}
\frac{\Gamma(s+j-1)}{\Gamma(s)j!}
\frac{g^{2(1-s)}\kappa^{1+2s}(1-r)^{2j}}{(1+r)^{1+2s+2j}}
\frac{1}{45(2s+2j-3)} \int d^2x
\nonumber\\
&\times&
\Big(w_0 (j,s) \frac{\zeta_{R}(2s-3)}{\kappa^4}
    +w_2 (j,s) \frac{\zeta_{R}(2s-1)}{\kappa^2}
    +w_4 (j,s)\zeta_{R}(2s+1)
\Big)\,,
\end{eqnarray}
where, as before, we must consider the $j=0$, $j=1$ and $j\geq 2$
modes separately. The contribution from the $j=0$ mode is the simplest
one, and
analytic continuation to $s= 0$ results in
\begin{eqnarray}
\frac{d}{ds}\Big(G(0,s)\Delta(0,s)\Big)\Big|_{s\to 0}
&=&
\frac{1}{2^4\pi}
 \frac{1}{135}
 \frac{g^2\kappa}{(1+r)}
\int d^2x
\Big\{
\Big(-\frac{1}{s-1}
     -\frac{2}{2s-3}
     +2\ln\Big(\frac{\kappa}{g(1+r)}\Big)
\Big)
\\
&\times&
\Big(
 w_0(0,0)\frac{\zeta_{R}(-3)}{\kappa^4}
+w_2(0,0)\frac{\zeta_{R}(-1)}{\kappa^2}
+w_4(0,s)\zeta_R (2s+1)\Big|_{s\to 0}
\Big)
\nonumber\\
&+&
 w_0(0,0)\frac{2\zeta_R{}'(-3)}{\kappa^4}
+w_2{}'(0,0)\frac{\zeta_R(-1)}{\kappa^2}
+2w_2(0,0)\frac{\zeta_R{}'(-1)}{\kappa^2}
\nonumber\\
&+& \partial_s \Big(w_4(0,s) \zeta_R (2s+1)\Big)\Big|_{s\to 0}
\Big\}
\nonumber\\
&=&
\frac{1}{2^4\pi}
 \frac{1}{135}
 \frac{g^2\kappa}{(1+r)}
\int d^2x
\Big\{
\Big(\frac{5}{3}
     +2\ln\Big(\frac{\kappa}{g(1+r)}\Big)
\Big)
\Big(
 3\frac{(1+r)^4}{\kappa^4}
+30\frac{(1+r)^2}{\kappa^2}
-42
\Big)
\nonumber\\
&& -50 \frac{(1+r)^2}{\kappa^2}
+ \frac{720(1+r)^4}{\kappa^4}\zeta_R{}'(-3)
-\frac{720(1+r)^2}{\kappa^2}\zeta_R{}' (-1)
+14(11-6\gamma)
\Big\}\,,
\nonumber
\end{eqnarray}
where we have used the fact that
\begin{eqnarray}
 w_4(0,s)\zeta_R(2s+1)=-42+14(11-6\gamma)s+O(s^2)\,.
\end{eqnarray}

Similar steps follow for the $j=1$ mode. By setting $j=1$ in
$G(j,s)\Delta(j,s)$, differentiating with respect to $s$, and
then taking the analytic continuation as $s\to 0$, we obtain
\begin{eqnarray}
\frac{d}{ds}\Big(G(1,s)\Delta(1,s)\Big)\Big|_{s\to 0}
&=&
-\frac{1}{2^4\pi}
 \frac{1}{45}
 \frac{g^2\kappa(1-r)^2}{(1+r)^3}
\int d^2x
\Big\{
\Big(2
     +2\ln\Big(\frac{\kappa}{g(1+r)}\Big)
\Big)
\nonumber\\
&\times&
\Big(
 3\frac{(1+r)^4}{\kappa^4}
-30\frac{(1+r)^4}{\kappa^2(1-r)^2}
-14
 +\frac{60(1+r)^2}{(1-r)^2}
\Big)
\nonumber\\
&&
 -10 \frac{(1+r)^2}{\kappa^2}
+ \frac{720(1+r)^4}{\kappa^4}\zeta_R{}'(-3)
+ \frac{720(1+r)^4}{\kappa^2(1-r)^2}\zeta_R{}' (-1)
\nonumber\\
&&-14
  -\frac{120(1+r)^2}{(1-r)^2}
  +\gamma\Big(-28
             +\frac{120(1+r)^2}{(1-r)^2} \Big)
\Big\}
\,,
\end{eqnarray}
where we have used the expansion around $s=0$:
\begin{eqnarray}
 w_4(1,s)\zeta_R(2s+1)
=-14+\frac{60(1+r)^2}{(1-r)^2}
+\Big[-14
  -\frac{120(1+r)^2}{(1-r)^2}
  +\gamma\Big(-28
             +\frac{120(1+r)^2}{(1-r)^2}
         \Big)
   \Big]s+O(s^2)\,.
\end{eqnarray}

For the modes with $j\geq 2$, we obtain
\begin{eqnarray}
&&\sum_{j=2}^{\infty}
 \frac{d}{ds}\Big(G(j,s)\Delta(j,s)\Big)\Big|_{s\to 0}
\nonumber\\
&=&
\frac{1}{2^4\pi}
 \sum_{j=2}^{\infty}
 \frac{1}{45j(j-1)(2j-3)}
 \frac{g^2\kappa(1-r)^{2j}}{(1+r)^{1+2j}}
\int d^2x
\nonumber\\
&\times&
\Big\{\Big[\psi(j-1)
     -\frac{2}{2j-3}
     +2\ln\Big(\frac{\kappa}{g(1+r)}\Big)
\Big]
\Big(
{w_4(j,s)\zeta_R (2s+1)\over \Gamma(s)}
\Big)
\Big|_{s\to 0}
\nonumber \\
&+&w_0(j,0)\frac{\zeta_R(-3)}{\kappa^4}
+w_2(j,0)\frac{\zeta_R(-1)}{\kappa^2}
+\frac{1}{2}w_4(j,0)
+\frac{1}{2}
\Big( 3\gamma w_4(j,0)
    +\partial_s w_4(j,s)\Big|_{s\to 0}
\Big)\Big\}
\nonumber\\
&=&
\frac{1}{2^4\pi}
 \sum_{j=2}^{\infty}
 \frac{1}{45j(j-1)(2j-3)}
 \frac{g^{2}\kappa (1-r)^{2j}}
      {(1+r)^{1+2j}}
\int d^2x
\nonumber\\
&\times&\Big\{
\Big[\psi(j-1)
     -\frac{2}{2j-3}
     +2\ln\Big(\frac{\kappa}{g(1+r)}\Big)
\Big]
\nonumber\\
&\times&2j(j-1)
\Big(7(-3+2j)(-1+2j)-\frac{30(-3+2j)(1+r)^2}{(1-r)^2}
+\frac{45(1+r)^4}{(1-r)^4}\Big)
\nonumber\\
&+&\frac{3(1+r)^4}{\kappa^4}
 +\frac{5(1+r)^2\big[6+4j^2-8j\frac{2+r(-1+2r)}{(1-r)^2}\big]}{\kappa^2}
\nonumber\\
&+&\frac{2}{(1-r)^4}
\Big[-21(1-r)^4
+84\gamma (1-r)^4 j^4
-16 j^3 (1-r)^2
\Big(-7(1-r)^2
     +9\gamma\big(3+r(-1+3r)\big)\Big)
\nonumber\\
&-&4j\Big(-2(1-r)^2(38+r(-1+38r))
     +9\gamma
     \big(13+r(8+r(18+r(8+13r))\big)
    \Big)
\nonumber\\
&+&12 j^2
  \Big(-(1-r)^2 \big(31+\gamma(-22+31r)\big)
+4\gamma\Big(17+\gamma(-8+r (27+r(-8+17r)))\Big)
  \Big)
\Big]\Big\}\,,
\end{eqnarray}
where in the first step we used that fact that
\beq
\left[\frac{w_4(j,s) \zeta(2s+1) }{\Gamma(s)} \right]' =
\frac{1}{2} w_4'(j,0) + \frac{3}{2}  \gamma w_4(j,0)\,.
\eeq

%%%%%%%%%%%%%%%%%%%%%%%%%%%%%%%%%%%%%%%%%%%%

\section{Contribution from the axisymmetric modes}

The $n=0$ mode, $\lambda_{m0}$, can be treated independently as follows.
Using the binomial expansion, we straightforwardly obtain
\begin{eqnarray}
\zeta_{0}(s)
&=&\int d^2x
\frac{(2g)^{2(1-s)}}{2\pi(s-1)}
 \sum_{m=1}^{\infty}
 \Big[(m+\frac{1}{2})^2-\frac{1}{4}\Big]^{1-s}
\nonumber\\
&=&\int d^2x \frac{(2g)^{2(1-s)}}{2\pi}
   \sum_{j=0}^{\infty}
   \frac{2^{-2j}\Gamma(j+s-1)}{j!\Gamma(s)}
  \Big[
   \zeta_{H}(2s+2j-2,\frac{1}{2})
 -\Big(\frac{1}{2}\Big)^{2(1-s-j)}
  \Big]
\,.
\end{eqnarray}
This expression requires no further regularization, because the
analytic continuation is already {\it built-in} to this type of zeta
function. The final result is
\begin{eqnarray}
\zeta_0{}'(0)
=
\int d^2 x\,
g^2\,\frac{\zeta_R{}(3)}{\pi^3},
\end{eqnarray}
where we have used the fact that
\begin{eqnarray}
\sum_{j=2}^{\infty}
\frac{2^{-2j}}{j(j-1)}
 \Big[\zeta_H(-2+2j,\frac{1}{2})
 -\big(\frac{1}{2}\big)^{2(1-j)}
 \Big]
=\frac{1}{4}
  \big( -1+\ln 2
  \big)
+\frac{7}{8\pi^2}
 \zeta_R(3)\,,
\end{eqnarray}
 and $\zeta_R'(-2)=- \zeta_R(3)/4\pi^2$
, which is derived from $\zeta_R(z)=\pi^{z-1/2} \zeta_R(1-z)
[\Gamma((1-z)/2)/\Gamma(z/2)]$. An important observation is that
(C2) does not agree with the corresponding expression obtained via
the WKB approximation, by a factor of $-3/8$ (see equation (D36) in
Ref.~\cite{mns}).

%%%%%%%%%%%%%%%%%%%%%%%%%%%%%%%%%

\end{document}